\documentclass[draft]{agujournal2019}
\usepackage{url} 
\usepackage{lineno}
\usepackage[inline]{trackchanges} 
\usepackage{soul}
\usepackage[T1]{fontenc}  
\usepackage[utf8]{inputenc}
\usepackage[english]{babel}

\usepackage[intlimits]{amsmath}

\usepackage{bm}

\usepackage{amssymb}
\usepackage{eurosym}

\usepackage{enumitem}
\setlist{nosep}

\usepackage[implicit=false]{hyperref}
\hypersetup{
    colorlinks=true,
    urlcolor=blue,
}

\newcommand{\bvec}[1]{\mathbf{#1}}
\newcommand{\pa}{\partial}

\draftfalse

\journalname{JGR: Space Physics}

\begin{document}

%
%


\title{Influence of Solar Sails on Magnetic Field Measurements in Space Plasmas}

%
%




\authors{Konstantinos~Horaites\affil{1,2}, Juan~V.~Rodriguez\affil{1,2}, Ying~Liu\affil{1,2}}

\affiliation{1}{CIRES, University of Colorado, Boulder, CO, USA}
\affiliation{2}{NOAA National Centers for Environmental Information, Boulder, CO, USA}






\correspondingauthor{Konstantinos Horaites}{konstantinos.horaites@colorado.edu}



\begin{keypoints}
\item We quantify how two effects at a solar sail, eddy currents and magnetic pileup, can perturb in-situ magnetic field measurements.
\item For a hypothetical but realistic mission with a 40$\times$40~m sail, the eddy current and magnetic pileup effects are negligible ($<$1\% error). 
\item Overall, these effects can become significant for larger sails, higher ambient plasma densities, and higher magnetometer sampling rates.
\end{keypoints}

%
%

%
%


\begin{abstract}
Solar sail technology is ready to be deployed in a satellite mission carrying a science-grade magnetometer. 
In preparation for such a mission, it is essential to characterize the interactions between the sail and the ambient plasma that could affect the magnetometer readings.
The solar wind magnetic field is a key parameter in space weather prediction, because it governs the energy-releasing magnetic reconnection process at Earth's magnetopause.
This paper investigates the influence of solar sails on the ambient magnetic field, particularly focusing on two critical electromagnetic effects: eddy currents and magnetic pileup.
We find the induced eddy currents in the metallic sail can significantly perturb the local magnetic field at high frequencies.
We also suggest that magnetic pileup can influence the spacecraft's environment when the sail size is comparable to the electron kinetic scales of the surrounding plasma.  This research provides an initial guide for determining when sail-plasma interactions could impact magnetometer performance.
\end{abstract}

\section*{Plain Language Summary}
A solar sail is a large thin film of reflective material that is used to gradually propel spacecraft through the nearly-frictionless environment of outer space. A solar sail is pushed by the Sun's light,
much like how a boat's sail is pushed by the Earth's winds. Despite their promise, solar sails have only been employed in a few satellite missions to-date, so that relatively little is known about them. One outstanding question concerns whether the metallic film that gives the sail its reflectivity can interact with the ambient space ``plasma'' of charged particles. Such interactions can affect satellite magnetometers, instruments that are used ubiquitously to measure the magnetic field near the spacecraft. We investigate two physical effects caused by the sail, eddy currents and electron-scale magnetic pileup, that could disturb magnetometer measurements. We estimate the strength of these effects, and use this information to develop guidelines for the allowable sail size and magnetometer sampling frequency. The results can be applied to inform satellite design and mission planning. 

%
%
%

%


%
%
%
%

\section{Introduction}

As described in \cite{barnes_2014_sunjammer, eastwood_2015_sunjammer_overview, lotoaniu23, johnson24_solar_sail_smallsat}, the emerging technology of solar sails could allow a satellite to be placed in a halo orbit situated between the Sun and the L1 Lagrange point. 
This is because the momentum imparted by the sail's passive reflection and absorption of sunlight would help counteract the pull of the Sun's gravity, shifting the equilibrium location sunward from L1. 
At such sub-L1 locations, 
the solar wind plasma could be continuously monitored to forecast Earth-bound space weather with improved lead time.  
The southward component of the incoming solar wind magnetic field is especially important for space weather, because this component modulates the magnetopause reconnection and the resulting transmission of energy from the solar wind to Earth's magnetosphere.

Solar sail missions have been actively considered in the United States, and have received government funding for research and development, but no major mission has yet flown. 
A prominent example from recent history is the Solar Cruiser mission \cite{cannella21_solar_cruiser_design, johnson22_nasa_solar_cruiser}, which was initially approved but ultimately canceled in 2022 by NASA due to budget and schedule constraints.
In 2014, the near-L1 Sunjammer mission \cite{barnes_2014_sunjammer} was canceled for similar reasons.
Building confidence in solar sails, by addressing remaining uncertainties surrounding their use, is needed to facilitate the adoption of this technology.

One such outstanding question, that should be considered in advance of any solar sail mission, concerns the ``potential interaction between the sail structure or its materials and the sensors [...]'' \cite{lotoaniu23} in the solar wind environment.
These interactions could potentially affect electromagnetic field measurements \cite{eastwood14}, which, alongside particle measurements, provide a fundamental characterization of a spacecraft's local plasma environment.
By the same token, if the sail-plasma interactions are shown to be negligible or at least mitigable, then agencies can move forward more confidently with designing a space weather monitor that uses a solar sail for its sub-L1 stationing.

Here we consider the physics relevant to onboard magnetic field measurements made by magnetometers, emphasizing how these effects depend on plasma time- and length-scales.
The estimates provided here are intended to serve as an {\it a priori} guide for determining the regime where the sail-plasma interactions are significant.
Such interactions would need to be accounted for as part of a rigorous ``magnetic cleanliness'' program \cite<e.g., >[]{acuna04} of a spacecraft outfitted with a solar sail.  
Throughout this document, we will study the plasma-sail interaction in the context of a hypothetical mission, whose relevant characteristics are borrowed from recent Solar Cruiser-like design concepts \cite{johnson22_nasa_solar_cruiser, lotoaniu23, johnson24_solar_sail_smallsat}.

This document describes two magnetic effects of an object immersed in a flowing plasma: eddy currents and electron-scale magnetic field pileup.
Eddy currents are a known issue in conducting spacecraft that can affect magnetometer measurements \cite<e.g., >[]{kotsiaros20_sc_eddy_currents_juno}, but have not been studied in the context of solar sails.
Magnetic pileup is a familiar process that occurs at global scales outside planetary magnetospheres, but also appears in small-scale contexts such as magnetic reconnection \cite{guan25}.
Eddy currents and magnetic pileup depend on the physical size of the object, and therefore become a concern due to the large dimensions of the lightweight metallic sails used in space~\cite{eastwood14}. 
We highlight these two effects in the context of solar sail measurements.
The following arguments consider the electromagnetic effects of the sail alone, neglecting the structural booms.
Namely, we assume the booms are made of low-conductivity, non-magnetic composites used in space applications \cite<e.g., >[]{arce19_juice_mag_boom, ma24_space_materials}, such as the Carbon Fiber
Reinforced Polymer (CFRP) that was planned for Solar Cruiser sail mission \cite{cannella21_solar_cruiser_design}.

This paper is organized as follows. In Section~\ref{param_sec}, we provide background information and estimate the physical parameters of the system. In Section~\ref{interaction_sec}, we quantify the effect of the spacecraft-sail interactions (see Appendix~\ref{appendix_sec} for additional details). We discuss our results in Section~\ref{discussion_sec}, and summarize key conclusions in Section~\ref{summary_sec}.

\section{Sail and Plasma Parameters}\label{param_sec}
To develop a clear discussion, we adopt representative values for the sail, magnetometer, and ambient solar wind plasma parameters. We first present a summary in Table~\ref{params_table}, below---the choices for these values are explained in the corresponding subsections that follow.

\begin{center}
\begin{table}[h!]
    \caption{Summary of sail, magnetometer, and solar wind plasma parameters. The solar wind parameters are derived from average measurements near the L1 Lagrange point~\cite{klein_2019_sw_1_au_parameters}. For each variable, its name, mathematical symbol, and a reference value are shown in the table. The reference values are used to estimate the significance of the sail-plasma interactions throughout this document.}
    \label{params_table}
    \begin{tabular}{l l|c|r} 
     & \textbf{Name} & \textbf{Symbol} & \textbf{Reference Value}\\
      \hline
      {\bf Sail}  & Side length & $L$ & $40.7~m$ \\
       & Substrate thickness & $H$ & $2.5 \times 10^{-6}~m$ \\
       & Coating thickness & $h$ & $1 \times 10^{-7}~m$ \\
       & Conductivity (Aluminum)& $\sigma_{Al}$ & $3.77 \times 10^7~S~m^{-1}$ \\  
       & Permeability & $\mu~(=\mu_0)$ & $4 \pi \times 10^{-7} N~A^{-2}$ \\
      \hline
       {\bf Magnetometer}  & Sampling frequency & $f_{mag}$ & $10~Hz$ \\
            & Error tolerance (worst-case) & $|\Delta B_i/B_i|$ & $<0.01$ \\
      \hline
       {\bf Solar wind}  & Number density & $n$ & $5.56 \times10^{6}~m^{-3}$\\
        & Flow speed & $v_{sw}$ & $4.13\times 10^5~m~s^{-1}$ \\ 
        & Magnetic Field & B & $5.07\times10^{-9}~T$ \\ 
        & Temperature & $T_p$, $T_e$ & $1.17\times 10^5~K$, $1.47\times 10^5~K$  \\
        & Larmor radius & $\lambda_p, \lambda_e$  & $9.0 \times 10^4~m$,  $2.4 \times 10^3~m$  \\
        & Inertial length & $\rho_{p}, \rho_{e}$  &  $9.6\times 10^4~m$,  $2.3 \times 10^3~m$ \\
        & Debye length & $\Lambda_D$  &  $11.2~m$
    \end{tabular}
\end{table}
  \end{center}



\subsection{Sail}

Solar sails are constructed from thin sheets of a substrate polymer (e.g., Mylar or Kapton), coated with a layer of reflective conducting metal---typically aluminum---some tens of nanometers thick \cite{hollerman03}.
Following the examples set by the Solar Cruiser mission and other modern solar sail designs, we assume the mission's sail to be constructed from a polyimide film with thickness $H=2.5\times10^{-6}~m$ \cite{johnson22_nasa_solar_cruiser, johnson24_solar_sail_smallsat}, coated with a layer of Vapor Deposited Aluminum (VDA) of thickness $h=1\times 10^{-7}~m$ \cite{johnson_2020_solar_cruiser_tech_report, davoyan_2021_materials_for_solar_sailing}.
Similarly, we assume a square sail with side length $L = 40.7~m$ \cite{lotoaniu23, johnson24_solar_sail_smallsat}.  
The coating can reflect electromagnetic radiation at frequencies below the plasma frequency of its free electrons---for aluminum, this cutoff frequency is in the ultraviolet regime.

More than 90\% of optical light (near the peak of the Sun's spectrum) is reflected by a $>$50 nm-thick aluminum sheet \cite{hasswaylonis61}, and the rest of the energy is mostly absorbed rather than transmitted. This is because for a normally-incident transverse electromagnetic wave, the strength of the transmitted signal falls off exponentially in the metal, with a characteristic scale known as the ``skin depth'' $\delta(f)$ of the material that depends on the frequency $f$: 

\begin{equation}\label{skin_depth_eq}
\delta(f) = \sqrt{\frac{1}{\pi \mu \sigma f}},
\end{equation}

\noindent where $\mu$ is the material's permeability and $\sigma$ is its conductivity. For non-magnetic materials such as aluminum and the sail's polymer substrate, we may take $\mu$ to approximately be the permeability of free space \cite<e.g., >[]{hasswaylonis61}. So we will assume for the entire sail that $\mu \approx \mu_0$. 
For aluminum's conductivity $\sigma_{Al} = 3.77 \times 10^7$ S/m, the skin depth of red light ($f_{red}$ = 4.3e14 Hz) is $\delta(f_{red})$~=~4~nm~$\ll h$, so virtually all of the unreflected optical light will be absorbed before it can pass through the coating.

We additionally note that most ($>95\%$) of the solar wind protons, which have typical flow speeds $v_{sw}\sim 400$~km/s (energy $\sim$keV), will be absorbed by the 100-nm sail coating \cite{sznajder_2020_solar_sail_blistering}. So we may expect a shortage of protons immediately downstream of the sail, which helps to form a spacecraft ``wake'' structure \cite<e.g., >[]{wang_hu_2018_e_fluid_approximation_plasma_wake}.


\subsection{Magnetometer}

We will assume the hypothetical mission carries a fluxgate magnetometer similar to that on the GOES-R spacecraft \cite{lotoaniu19}.
Such a magnetometer could sample the vector magnetic field $\bvec{B}$ at a cadence $f_{mag}=10$~Hz (on GOES, low-pass filtered to $\sim$2.5~Hz). 
We base our discussion on this magnetometer for two reasons. 
First, fluxgate magnetometers are the most popular instruments for measuring the vector magnetic field in space applications \cite{acuna02}.
Secondly, an actual spare magnetometer from the GOES mission was argued to be appropriate for a near-L1 solar wind monitor in a white paper \cite{lotoaniu23}. 

In this study, we must also estimate the instrumental accuracy required for a solar sail mission.
Referring to the abovementioned white paper, the GOES spare magnetometer would meet NOAA's ``threshold'' operational requirements for an L1 spacecraft, as defined in the NOAA Space Weather Mission Service Area Consolidated Observational User Requirements List (COURL). 
The COURL document stipulates an error tolerance of $\pm 1$~nT (per axis) for fields $<$100~nT, and 1\% tolerance for fields $>$100~nT. 
For simplicity, in this analysis we will impose a more stringent ``worst-case'' requirement on the relative error tolerance, $|\Delta B_i/B_i| < 0.01$, for each magnetic field component $B_i$ \cite{lotoaniu23}.
This simplified criterion automatically meets the COURL threshold, and is more easily applied to the wide range of conditions present in the variable solar wind.

Besides being directly applicable to the magnetometer's instrumental accuracy, the magnetic error tolerance requirements also effectively limit the perturbation fields that can be allowed in the observed plasma environment.  
Therefore, in the present analysis we will interpret this same threshold to apply to the strength of the perturbation field $\bvec{B^\prime}$ that arises from the plasma-sail interaction, as compared to the background field $\bvec{B}$---i.e., we require $|B_i^\prime /B_i| < 0.01$.

Commonly, the magnetic effect of spacecraft-plasma interactions are mitigated by placing the magnetometer on a long boom \cite{acuna04}. This is because 
``most interference fields are dipole in nature and vary as $1/r^3$, where $r$ is the distance between the noise source and the sensor'' \cite{hospodarsky25}. The dual magnetometer technique \cite<e.g., >[]{acuna02} can be applied to correct the measurements in this dipole regime. However, this works only for fields far from the source object; localized (multipolar) fields may still be expected at scales comparable to the object size.
So, any currents in the sail should produce significant fields at scales comparable its side-length $L$ (Table~\ref{params_table})---even Voyager's record-length 13-meter boom would not extend beyond these fields. It is then especially important to understand the local magnetic fields caused by the solar sail, because a realistic boom cannot outreach them.

\subsection{Solar Wind}

The solar wind plasma at L1 has been thoroughly characterized, by in situ spacecraft measurements spanning multiple solar cycles. As shown in Table~\ref{params_table}) above, we adopt the average density,  magnetic field magnitude, and temperature as given in Table~1 of \cite{klein_2019_sw_1_au_parameters}: $n=5.56~\times 10^6~m^{-3}$, $B = 5.07\times10^{-9}~T$, $T_e=1.47\times 10^5$~K, $T_p=1.17 \times 10^5~K$ (subscripts $e$ and $p$ denote electron and proton, respectively). The slow solar wind ions and electrons flow together radially with an average bulk speeds $v_{sw} = 413$~km/s. Relevant to the present discussion of kinetic-scale plasma phenomena, we also introduce the Larmor radius of thermal particles $\lambda_j = \sqrt{2 m_j T_j}/e B$ and the inertial length $\rho_j = \sqrt{m_j c^2/4 \pi n e^2}$, where $e$ is the elementary charge, $c$ is the speed of light, and $m_j$ is the mass of particle species $j$. We note that the electron scales $\lambda_e$, $\rho_e$ are roughly 2--3~km, 
while the proton scales are larger by factor equal to the square root of the proton-electron mass ratio: $\lambda_p$, $\rho_p$$\sim$100~km. Indeed, the fact that kinetic scales are similar for a given species, $\lambda_j\sim \rho_j$, follows directly from the fact that in the solar wind the plasma beta is of order unity \cite<e.g., >[]{boldyrev13}.  The electron Debye length $\Lambda_D$, which governs the size of electrostatic structures in the plasma, is given by $\Lambda_D =\sqrt{\epsilon_0 k_B T_e / n e^2}$, where $\epsilon_0$ is the permittivity of free space and $k_B$ is Boltzmann's constant. In the solar wind, typical values are $\Lambda_D \sim 10~m$.

\begin{figure}
\includegraphics[width=\linewidth]{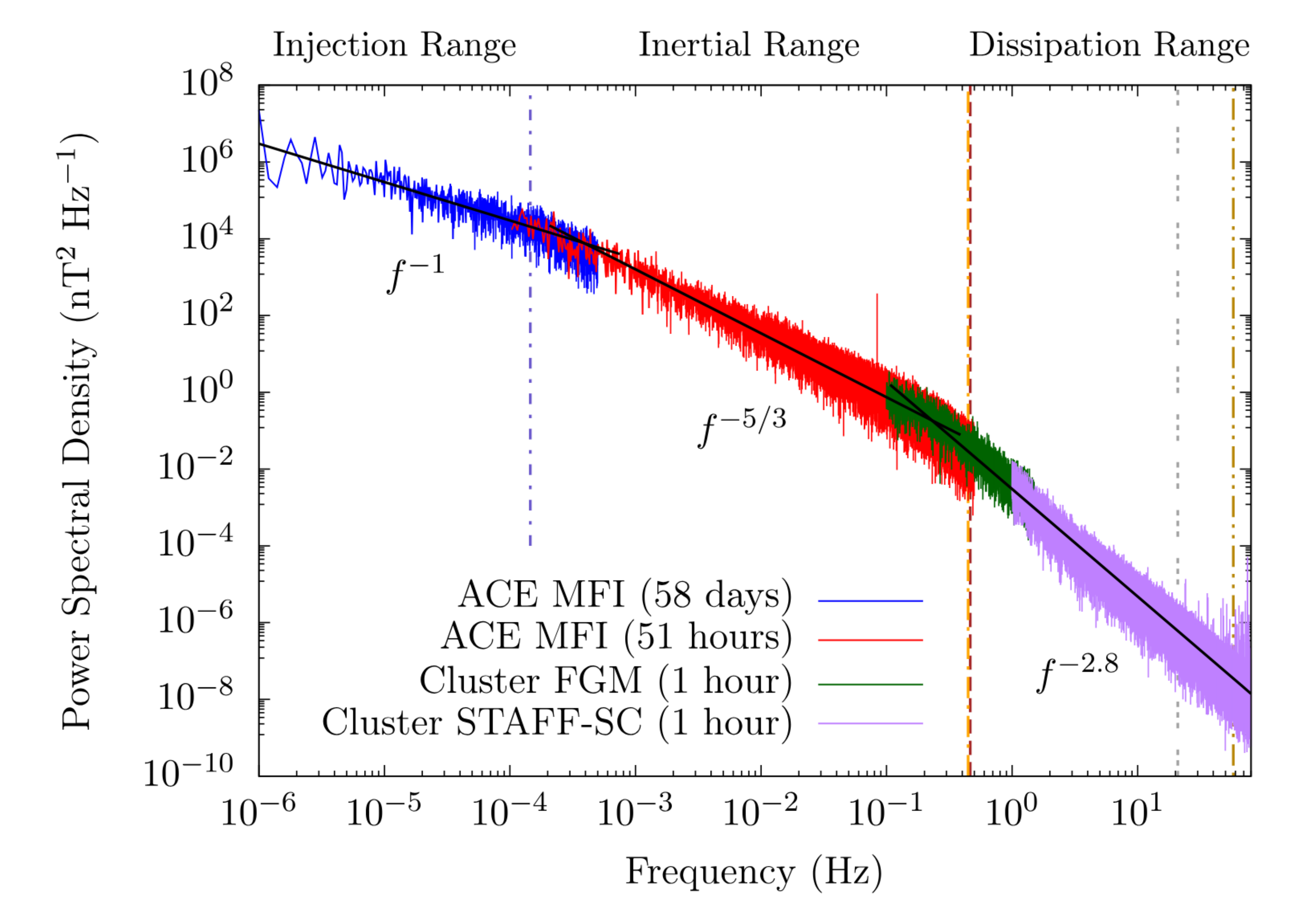}
\caption{The magnetic frequency power spectrum in the solar wind, modified with author's permission from \cite{verscharen19}. 
The plot combines magnetometer data from the ACE and Cluster missions, to display the spectrum over a broad range of frequencies.
Note that the assumed magnetometer's Nyquist frequency  $f_{mag}/2 = 5~Hz$  lies in the dissipation range. The instrument covers the lower-frequency (injection and inertial) ranges of the turbulent spectrum as well. }
\label{freq_spec_fig}
\end{figure}

The magnetic fields observed by the magnetometer are highly variable, 
so that the solar wind sustains a spectrum of fluctuations at frequencies $\leq 5$~Hz (i.e. at frequencies $\leq f_{mag}/2$~Hz accessible to the magnetometer by Nyquist's theorem); see Figure~\ref{freq_spec_fig}.
We note that light waves are strongly damped at these magnetometer-relevant frequencies, so the electromagnetic fluctuations in this regime are rather attributed to plasma waves. Although linear waves do exist in the solar wind, as they can be generated by kinetic instabilities, in the typical paradigm the observed magnetic fluctuations arise from turbulent dynamics. A representative magnetic power spectrum is shown in Figure~\ref{freq_spec_fig}. The fluctuations are driven by the deposition of energy at large scales, known as the injection range, which initiates a cascade of energy to successively smaller scales. The fluctuations become highly organized in the turbulence's so-called inertial range, typified by its power spectral index -5/3. The inertial range fluctuations are widely agreed to be parallel-propagating, nonlinearly-interacting Alfv\'en waves. In the dissipation range $\gtrsim$~1~Hz, which has a steeper spectral index of approximately -2.8, it has been argued that most of the fluctuation power is supplied by kinetic Alfv\'en waves \cite{boldyrev13}. However, other small-scale waves such as whistler waves may also be observed in some proportion.

\section{Sail-Plasma Interaction}\label{interaction_sec}

In this section, we analyze the sail-plasma interaction's effect on the local magnetic field. In Section~\ref{diffusion_sec}, we demonstrate that magnetic fields in the solar wind will rapidly permeate the thin material of a solar sail. This justifies the quasistatic approximation in Section~\ref{eddy_sec}, where we calculate the perturbation magnetic field $\bvec{B^\prime}$ generated by induced eddy currents in an idealized sail. In Section~\ref{pileup_sec}, we argue that the effect of magnetic field pileup can perturb the B-field when the sail size $L$ exceeds the electron inertial length.

\subsection{Magnetic Diffusion}\label{diffusion_sec}

The frequencies $f \leq f_{mag}$ relevant to the hypothetical mission's magnetometer are slow: electromagnetic signals can cross the aluminum coating thickness $h$ much faster than the relevant time scale $1/f_{mag}$. In such a regime, i.e. $f_{mag} \ll c/h$, we may neglect the displacement current in Maxwell's equations. In this so-called ``(magnetic) quasistatic approximation'', magnetic fields diffuse through matter. Specifically,  the space- and time-dependent magnetic field $\bvec{B}(\bvec{x}, t)$ obeys a diffusion equation:
 
\begin{equation}\label{mag_diff_eq}
\frac{\partial \bvec{B}}{\partial t} = \frac{1}{\mu \sigma} \nabla^2 \bvec{B},
\end{equation}

\noindent where $\mu$ and $\sigma$ are, respectively, the permeability and conductivity of the material. 

Let us estimate the distance $D$ over which magnetic fields diffuse in a time scale $\tau$. From (\ref{mag_diff_eq}), we use the approximations (ignoring factors of order unity) $\partial/\partial t \sim 1/\tau$ and $\nabla^2 \sim 1/D^2$ to find:

\begin{equation}\label{Delta_eq}
D \sim \sqrt{\frac{\tau}{\mu \sigma}} \sim \delta(1/\tau).
\end{equation}

\noindent For a signal with frequency $f$, the characteristic time is $\tau = 1/f$, and we recognize the spatial scale of diffusion is simply the skin depth of the material: $D \sim \delta(f)$. For signals at magnetometer-relevant frequencies $f \lesssim f_{mag}$ in aluminum (conductivity $\sigma = \sigma_{Al}$), we find $D \gtrsim \delta(f_{mag}) = 2.6 $ cm. That is, 
$D \gg h$, and the observable magnetic field diffuses right through the metal without significant attenuation. Such a conclusion is supported by generally accepted knowledge, namely that aluminum is ineffective in attenuating applied low-frequency magnetic fields \cite<e.g., >[]{hoburg95}.


We finally consider the magnetic effect of the sail's polymer substrate, which we have so far neglected. As an insulator, the substrate is not easily described by the theory of magnetic diffusion. But, since its material is non-magnetic (i.e. $\mu$ is just the permeability of free space, see Table \ref{params_table}), we may conclude the substrate also does not affect the applied fields.
So, we will assume the applied low-frequency fields of the solar wind are homogeneous across the sail membrane. 

\subsection{Eddy Currents}\label{eddy_sec}

Even in the regime $D \gg h$, however, ``very significant shielding can be accomplished [...] with materials such as copper and aluminum, when current loops that circulate around the large scale dimensions of the shield result in an induced flux density [...]'' \cite{hoburg95}. In other words, even if the magnetic field $\bvec{B}(\bvec{x}, t)$ diffuses efficiently through the sail membrane, 
the time-variation $\pa \bvec{B}/\pa t$ in the material may induce large-scale circulating currents that help to screen out the same field. This is evident from the familiar Lenz's law, which states that induced currents tend to counteract the variation of the applied magnetic field. In the following, we model an idealized interaction between the sail and the solar wind plasma fields, explicitly calculating the induced large-scale eddy currents and the resulting magnetic perturbation.

\begin{figure}
\includegraphics[width=1.0\linewidth]{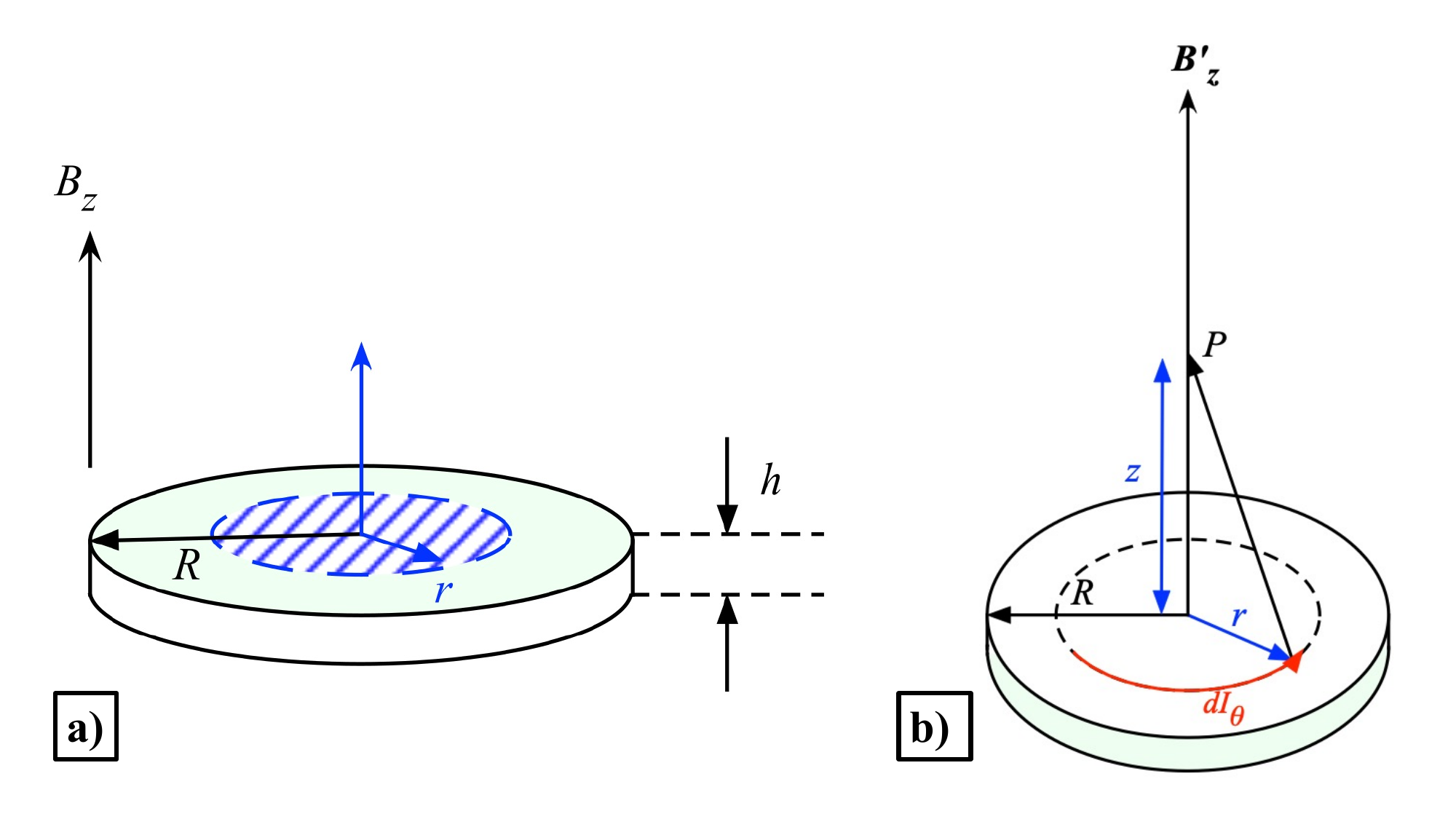}
\caption{We model the solar sail as a thin circular disk, to estimate the effects on the square sail of similar dimensions 
(figures from \cite{tan14_fermilab_eddy_currents}, modified and included with author's permission). {\bf a)} Diagram of a metal disk with radius $R$ and thickness $h$ suffused by a homogeneous time-varying magnetic field $\bvec{B}(\bvec{x}, t) = \bvec{\hat z}B_z(t)$. {\bf b)}  At an arbitrary point P along the axis of symmetry (cylindrical radius $r = 0$), we may calculate the perturbation field $\bvec{B_z^\prime}(\bvec{x}, t)|_{r=0} = \bvec{\hat z} B_z^\prime(z, t)$. The field $\bvec{B^\prime}$ is generated by the induced azimuthal current $dI_\theta(r)$.}
\label{tan_fig2}
\end{figure}

For simplicity, let's approximate the square $L\times L$ sail by a circular disk with radius $R=L/2$, centered on the $z$-axis as pictured in Figure~\ref{tan_fig2}a. The time-varying normal component of the field is responsible for eddy currents. So, we also assume the sail is in the presence of a homogeneous time-varying field $\bvec{B}(\bvec{x}, t) = \bvec{\hat z} B_z(t)$, and the skin depth greatly exceeds the sail's thickness: $\delta \gg h$. Accordingly, we assume this applied field $\bvec{B}(\bvec{x}, t)$ is unattenuated in the sail (see section~\ref{diffusion_sec}), but its time-variation induces a large-scale current that produces a perturbation field $\bvec{B^\prime}(\bvec{x}, t)$. We also assume $|\bvec{B^\prime}|/|\bvec{B}| \ll 1$ and $|\pa \bvec{B^\prime} /\pa t|/|\pa \bvec{B}/\pa t| \ll 1$  in the disk, so we need only consider the $\partial  \bvec{B} / \partial t$ term in Faraday's Law and may neglect the perturbation field term (the ``thin-sheet approximation'' \cite{knoepfel_2008_magnetic_fields_practical_use}). 
Noting the system is symmetric with rotations about the $z$-axis, we adopt the standard cylindrical coordinate system $\{z, r, \theta\}$.

Conveniently, this exact physics problem has already been worked out in detail in a Fermilab report (Section A, \cite{tan14_fermilab_eddy_currents}), albeit for an unrelated laboratory physics application. We rederive this result in Appendix~\ref{appendix_sec}; though our approach is more general in that it can predict the induced field over all space (eq.~\ref{B_rings_general_eq}). As a result of the cylindrical symmetry, an azimuthal current $dI_\theta(r)$ is induced in the disk that is proportional to the time rate-of-change of the applied field $\dot B_z(t)$. The induced current produces a perturbation field $\bvec{B^\prime(\bvec{x}, t)}$ that is purely axial along the axis of symmetry ($z$), i.e. $\bvec{B^\prime}(\bvec{x}, t)\Big|_{r=0} = \bvec{\hat z} B_z^\prime(z, t)$; see Figure~\ref{tan_fig2}. From the Biot-Savart Law, this perturbation field on the $z$-axis is:

\begin{equation}
\tag{\ref{B_rings_zaxis_eq}}
B_z^\prime(z, t) = -\frac{\mu_0 \sigma h}{4} \dot B_z(t) \frac{(|z|
- \sqrt{R^2 +z^2})^2}{\sqrt{R^2 +z^2}},
\end{equation}

\noindent Naturally, if the sail is coated on both sides with identical metallic layers, then the right side of (\ref{B_rings_zaxis_eq}) should be multiplied by 2.

Far from the spacecraft, in the regime $\alpha = R^2 / z^2 \ll 1$, we may expand (\ref{B_rings_zaxis_eq}) with respect to $\alpha$ to confirm the magnetic field exhibits a dipole ($\sim |z|^{-3}$) dependence:

\begin{equation}\label{Bz_prime}
B_z^\prime(z, t) \approx -\frac{\mu_0 \sigma h R^4}{16}\dot B_z(t) |z|^{-3} \text{;\ \ \ \  if $\alpha \ll 1$.}
\end{equation}

\noindent But, the behavior differs from a dipole close to the sail, in the regime $\alpha \gtrsim 1$. At the sail's exact center ($z=0$), where $|B_z^\prime(z,t)|$ maximal, the perturbation magnetic field is:

\begin{equation}\label{Bz_prime_z0_eq}
B_z^\prime(z=0, t) = -\frac{\mu_0 \sigma h R}{4} \dot B_z(t).
\end{equation}

Bearing in mind the broad spectrum of solar wind magnetic fluctuations (Fig.~\ref{freq_spec_fig}), let us consider a sinusoidal applied signal with frequency $f$: $B_z(t) \sim \exp(i 2 \pi f t)$. As is conventional, we treat each harmonically-varying quantity as complex (and the real part is physical). The ratio between the applied and generated magnetic field is:

\begin{equation}\label{eddy_current_B_eq_w_phase}
\frac{B_z^\prime(z=0, t)}{B_z(t)} = -i \gamma(hR, f).
\end{equation}

\noindent where we introduced the dimensionless variable $\gamma(hR, f) = \pi \mu_0 \sigma h R f /2$.

\begin{figure}
\includegraphics[width=0.5\linewidth]{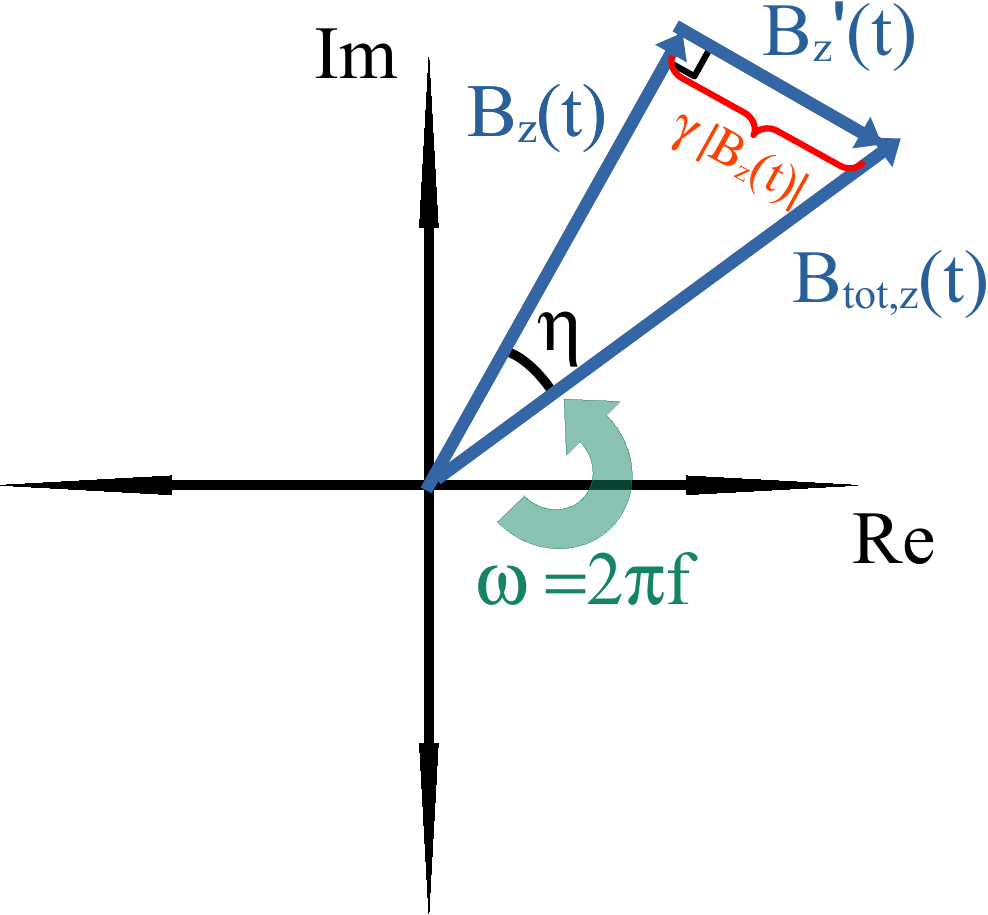}
\caption{The quantities $B_z(t)$, $B_z^\prime(z=0, t)$, and $B_{tot,z}(z=0, t)$ are assumed to be harmonically varying with angular frequency $\omega=2\pi f$, are shown in the complex plane---see eqs.~(\ref{eddy_current_B_eq_w_phase}) and (\ref{Btot_eq}).}
\label{B_Btot_complex_plane_fig}
\end{figure}

The factor of $-i$ in eq.~
(\ref{eddy_current_B_eq_w_phase}) represents  a 90-degree phase shift between the applied signal $B_z(t)$ and the induced field $B^\prime_z(z,t)$. As a result, the total field on the z-axis $B_{tot}(z, t)$, evaluated at the origin $z=0$, is larger in magnitude than $B_z(t)$, by a factor $\sqrt{1+\gamma^2}$:

\begin{equation}\label{Btot_eq}
\begin{split}
B_{tot,z}(z=0, t) &= B_z(t) + B^\prime_z(z=0, t)\\
&= (1 - i \gamma)\ B_z(t) \\
&=  B_z(t)\sqrt{1+ \gamma^2} \exp(i \eta),
\end{split}
\end{equation}

\noindent  and $\eta = \tan^{-1}(-\gamma)$ describes the relative phase. The magnitude and phase relationships between the harmonically-varying quantities $B_z(t)$, $B_z^\prime(z=0,t)$, and $B_{tot,z}(z=0, t)$ are shown in the complex plane in Figure~\ref{B_Btot_complex_plane_fig}.

Because of the 90-degree phase shift between $B_z(t)$ and $B_z^\prime(t)$, the two signals do not exhibit exact destructive interference. To quantify the relative error introduced by the induced signal, we consider two metrics: 1)  the maximum absolute difference between $\text{Re}(B_z(t))$ and $\text{Re}(B_{tot,z}(z=0,t))$, which is simply the magnitude $|B_z^\prime(z=0,t)|=const.$ and 2) the difference between the oscillation amplitudes $|B_z(t)|$ and $|B_{tot,z}(z=0,t)|$. Normalizing by the applied signal amplitude $|B_z(t)|$, we respectively define these two error metrics $\mathcal{E}_1>0$, $\mathcal{E}_2>0$ as follows:

\begin{equation}\label{eddy_current_B_err_eq}
\mathcal{E}_1=\frac{|B_z^\prime(z=0,t)|}{|B_z(t)|} = \gamma(hR, f) = \frac{\pi \mu_0 \sigma}{2} h R f,
\end{equation}

\begin{equation}\label{eddy_current_B_err_eq2}
\mathcal{E}_2=\frac{|B_{tot,z}(z=0,t)| - |B_z(t)|}{|B_z(t)|} = \sqrt{1+\gamma^2}-1.
\end{equation}

\noindent The error metric $\mathcal{E}_1$ defined in eq.~(\ref{eddy_current_B_err_eq}) is more relevant to single-time measurements of a sinusoidal signal, whereas the error metric $\mathcal{E}_2$ in eq.~(\ref{eddy_current_B_err_eq2}) is more relevant to the spectral amplitude at frequency $f$. To explain this, consider applying a narrow bandpass filter around the frequency $f$ to both the ambient solar wind magnetic spectrum and the magnetometer-measured spectrum, to respectively produce two sinusoidal signals $B_z(t)$ and $B_{tot,z}(z=0,t)$. Then $\mathcal{E}_1$ would describe the maximum time-dependent difference between the two filtered signals, and $\mathcal{E}_2$ would describe the difference in amplitudes of the two signals---in either case, normalized by $|B_z(t)|$. These error metrics increase with the field perturbation $B^\prime(z=0,t)$, and the case of no perturbation corresponds with $\mathcal{E}_1 = \mathcal{E}_2 = 0$.

Substituting the values from Table~\ref{params_table} into equation (\ref{eddy_current_B_err_eq}), we find that for a sail of radius $R=L/2$ with one-sided aluminum coating ($\sigma=\sigma_{Al}$) of thickness $h$, at magnetometer-relevant frequencies $f < f_{mag}/2$ the ratio is: 
$\mathcal{E}_1 < 7.6 \times 10^{-4}$.
That is, the anomalous magnetic field $B_z^\prime$ generated by the eddy currents is negligible compared to the background $B_z$ and within the required 1\% tolerance (Table~\ref{params_table}). Because of the proportionality 
$\mathcal{E}_1 \propto R$
in eq.~(\ref{eddy_current_B_err_eq}), large-scale eddy currents in average conditions may have a significant effect for larger sails, i.e. with radii $R \gtrsim 269$~m. 
The proportionality 
$\mathcal{E}_1 \propto f$
in eq.~(\ref{eddy_current_B_err_eq}) further implies the measurement error introduced by the induced field $\bvec{B^\prime}$ is more impactful for the high-frequency, small-scale features in the solar wind. 

\begin{figure}
\includegraphics[width=\linewidth]{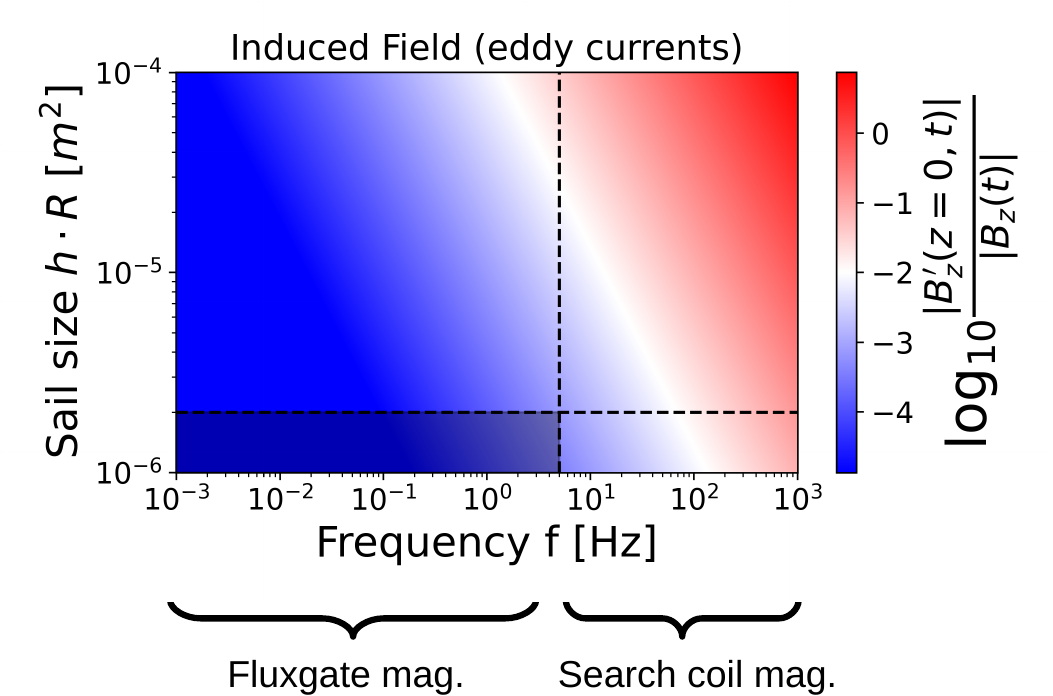}
\caption{The plot shows the magnetic error metric $\mathcal{E}_1 = |B^\prime_z(z=0,t)/B_z(t)|$,
from equation~(\ref{eddy_current_B_err_eq}). According to this formula, $\mathcal{E}_1$ depends linearly on both the frequency $f$ of the applied sinusoidal field and the sail dimension $h \cdot R$. The white diagonal line divides the plot, into a blue region that satisfies the worst-case error tolerance requirement $|B^\prime_z/B_z|<0.01$ and a red region that violates this requirement. The shaded region in the lower left corner is relevant to the hypothetical solar sail mission, as it is bounded by horizontal and vertical dashed lines which respectively show the mission's parameters $h \cdot R$ (where $R = L/2$) and Nyquist frequency $f_{mag}/2$ (Table~\ref{params_table}). The x-axis annotations show the approximate frequency regimes typical for fluxgate ($\lesssim$10~Hz) and search coil ($\sim$Hz--kHz) magnetometers.}
\label{sail_induced_B_fig}
\end{figure}

The bilinear dependence (eq.~\ref{eddy_current_B_err_eq}) of $\mathcal{E}_1$ on the frequency $f$ and sail dimension $h \cdot R$ (units $m^2$) is shown in Figure~\ref{sail_induced_B_fig}.  The shaded region in the figure, which is relevant to the hypothetical mission, falls within the blue region of the plot (which satisfies the error tolerance requirement 
$\mathcal{E}_1<0.01$).
The red region, correspondingly, violates this requirement. We note from the log-scale color bar that the far upper-right corner of the red region indicates an error $\mathcal{E}_1 = |B_z^\prime|/|B_z|>1$), which violates the quasistatic assumption of small perturbations (section \ref{eddy_sec}) used to derive the error. For large induced fields $|B_z^\prime| \gtrsim|B_z|$, this approximation breaks down, but naturally in this regime the induced field will still have a significant spurious effect. So although the quasistatic approximation doesn't hold for large perturbations, it is accurate at determining the white boundary (1\% error tolerance threshold) that separates the admissible (blue) and inadmissible (red) parameter regimes.

\begin{figure}
\includegraphics[width=\linewidth]{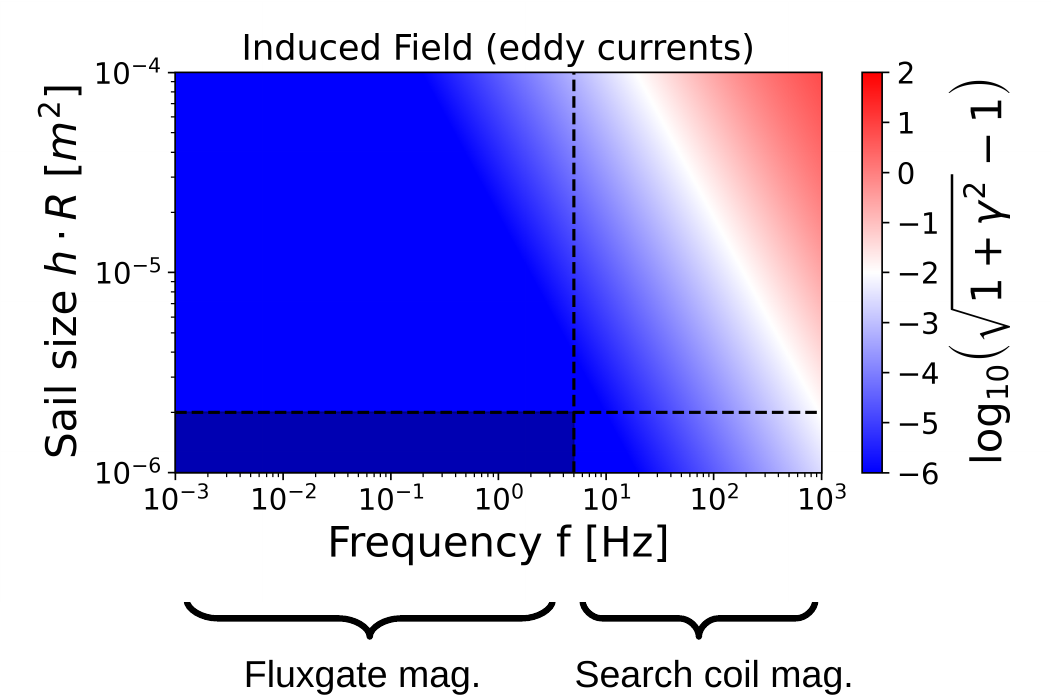}
\caption{The plot, analogous to Figure~\ref{sail_induced_B_fig}, shows the dependence of the magnetic error metric $\mathcal{E}_2$ (eq.~\ref{eddy_current_B_err_eq2}) on the frequency $f$ and sail dimension $h \cdot R$.}
\label{sail_induced_Btot_fig}
\end{figure}

In Figure~\ref{sail_induced_Btot_fig}, we plot the error metric $\mathcal{E}_2$ (eq.~\ref{eddy_current_B_err_eq2}) as a function of the frequency $f$ and sail dimension $h \cdot R$. 
Analogously to  Figure~\ref{sail_induced_B_fig}, the color scale divides the plot into a blue region $\mathcal{E}_2<0.01$ that meets our error tolerance requirement and a red region $\mathcal{E}_2 > 0.01$ that violates this requirement. 
The red region in Figure~\ref{sail_induced_Btot_fig} is smaller than in Figure~\ref{sail_induced_B_fig}, so that broadly speaking the metric $\mathcal{E}_1$ imposes a more stringent limitation than $\mathcal{E}_2$.
However, the need for either $\mathcal{E}_1$ or $\mathcal{E}_2$ to satisfy a $\Delta B_i/B_i$ threshold will depend on the science priorities of the spacecraft mission---respectively, accurate time series or spectral amplitudes---in the narrow-band range around frequency $f$.

Figures~\ref{sail_induced_B_fig} and \ref{sail_induced_Btot_fig} are relevant to the scalability of solar sail technology to magnetometer-carrying spacecraft missions, as science or operational considerations of future missions may push designs towards larger sails or higher-frequency sampling. 
Depending on the sail size, we observe that the perturbation field could violate our error tolerance requirements within the turbulent dissipation range ($f\gtrsim$Hz) of the magnetic power spectrum, and especially within the high-frequency regime ($\sim$kHz) sampled by search coil-type magnetometers.
Search coil magnetometers are especially used for studying wave phenomena in scientific contexts~\cite{hospodarsky_2016_search_coil_mags_review}.
An estimate of the range of allowable sail sizes, for a magnetometer with sampling frequency $f_{mag}$, is given by equation \ref{f_mag_threshold_eq} in section~\ref{summary_sec}.

Finally, we note that despite the solar wind's broad spectrum of fluctuations, the sampled vector magnetic field is actually fairly stable over short times. This is because at short time scales the sampled magnetic field's spectral amplitude 
is relatively weak (see Fig.~\ref{freq_spec_fig}). Rather than applying our single-frequency results to estimate the error within the solar wind's rich spectrum of fluctuations, for single-time measurements integrated over a short interval $\Delta t$ it is more direct to consider the total variation $\delta \bvec{B}(t, \Delta t) = \bvec{B}(t+\Delta t) - \bvec{B}(t)$, which has been studied empirically.  Introducing $B_0(t, \Delta t)$ as the average magnetic field magnitude over the time period $\tau = [t,t+ \Delta t]$, we may extrapolate Figure~3 of \cite{matteini18_1_f_scaling_and_compressibility} to infer that the typical normalized variation in the turbulent magnetic field is $|\delta \bvec{B}(t, \Delta t)|/B_0(t, \Delta t)$~$\sim 0.01 $ for time scales $\Delta t \sim 1/f_{mag}$ \cite{matteini18_1_f_scaling_and_compressibility}. If we decompose the solar wind magnetic field $\bvec{B}$ into an average and fluctuating part $\bvec{B}(\bvec{x}, t) = \bvec{B_{ave}}(\bvec{x}, t) +  \bvec{\tilde B}(\bvec{x}, t)$ over such a time range $[t, t+\Delta t]$, it follows immediately that $|\bvec{\tilde B}(\bvec{x}, t)| / |\bvec{B_{ave}}(\bvec{x}, t)| \sim 0.01$. And from the above arguments the solar wind fluctuating field $\bvec{\tilde B}(\bvec{x}, t)$ will typically be larger than the perturbation field $\bvec{B^\prime}$ it induces via the plasma-sail interactions. So, we may expect an ordering $|\bvec{B^\prime}| \ll |\bvec{\tilde B}| \ll |\bvec{B_{ave}|}$, suggesting that for single-time measurements of the magnetic field $\bvec{B}$ our observational requirement $|\Delta B_i/B_i| = |B^\prime_i / (\tilde B_i+ B_{ave,i})| < 0.01$ (Table \ref{params_table}) should be satisfied under typical conditions. 

\subsection{Magnetic Pileup}\label{pileup_sec}





In magnetohydrodynamics (MHD), the frozen-in magnetic field lines can become compressed when the plasma decelerates while approaching an obstacle.  This phenomenon is sometimes referred to as ``magnetic pileup'', and occurs for example in the so-named Magnetic Pileup Region (MPR) that separates Mars's magnetosheath from its induced magnetosphere \cite<e.g., >[]{acuna98}. At Mars's MPR, as well as the MPRs found at comets and at Venus, the magnetic pressure ($\propto B^2$) increases by a factor of 4-10 \cite{bertucci_mpb_mars_venus}. If large enough, a solar sail might also obstruct the particle flow and divert the frozen-in field.
Although the phenomenon is best-known in the context of single-fluid ideal MHD, valid at scales $>\rho_p$, magnetic pileup can still occur for a square sail whose side length $L$ falls in an intermediate ``electron scale'' regime $\rho_e < L< \rho_p$.
For such a mission, the effect of magnetic pileup could then become a source of error in magnetometer measurements.


A square solar sail with a scale-length $L\sim$40~m, as for the hypothetical mission, lies in the fully kinetic regime $L < \rho_e$. Therefore, the solar wind particles will not behave like a fluid as they pass by such a sail, and no magnetic pileup should be expected. However, it is feasible that the regime $L\sim \rho_e$ could be approached by future missions, through a combination of pushing the technology to larger sail sizes and by sampling the denser plasma environment near the Sun (because $\rho_e \propto 1/\sqrt{n}$). Indeed, these factors may coincide because larger sails allow for sub-L1 stationing closer to the Sun \cite{lotoaniu23}. Larger sails may also accommodate higher-mass payloads, which is desirable for incorporating more scientific instruments. Transient events such as coronal mass ejections may also help to produce high-density conditions where the impact of pileup becomes relevant to magnetometer. For example, in \cite{liu_2014_stereo_extreme_cme} an extreme CME was reported with a peak number density $\gtrsim100 cm^{-3}$, which would reduce the inertial length $\rho_e$ by a factor of $\sim 5$ compared to ambient conditions, to $\rho_e \approx 500$~m.

The mechanism for electron-scale pileup is analogous to the ideal MHD case. At subproton scales ($<\lambda_p, \rho_p$), the protons form a neutralizing background and the current is carried mostly by the electrons. The appropriate set of governing equations is highly dependent on the physical context, with variations that include Hall MHD \cite{huba23_hall_mhd, toth08} and electron MHD \cite{gordeev94_emhd, lyutikov13_emhd_turbulence} (which may be seen as a small-scale limit of Hall MHD, see \cite{cholazarian09_emhd_turbulence_simulations}). Other related models have been developed to describe electron-scale kinetic Alfv\'en turbulence \cite{boldyrev13} and the Kelvin-Helmholtz instability \cite{tsiklauri24}. A key similarity of these models is that the magnetic field is frozen into the {\it electron} fluid \cite{toth08, lyutikov13_emhd_turbulence, boldyrev13}. Much like in ideal MHD, accelerations of this electron fluid will in general affect the frozen-in field. This means that magnetic pileup can occur at electron scales, as it has been observed for example in particle-in-cell simulations of reconnection in an electron-scale current sheet \cite{guan25}.  

Even in the electron-fluid regime $L \gtrsim \rho_e$, the electrons need to see the sail as an obstacle in order to be diverted around it and cause magnetic pileup. For example, the conducting ionospheres of Venus and Mars \cite{zhang91_venus_magnetic_barrier, ho02_radio_prop_mars_handbook} and of comets \cite{coates_1997_cometary_environment_review} can provide such obstacles to the solar wind. But a solar sail's conductive coating is too thin to efficiently exclude the incoming magnetic field in this manner. 
Rather, magnetic diffusion dominates because $D \gg h$ in the VDA (Section~\ref{diffusion_sec}). 

\begin{figure}
\includegraphics[width=0.8\linewidth]{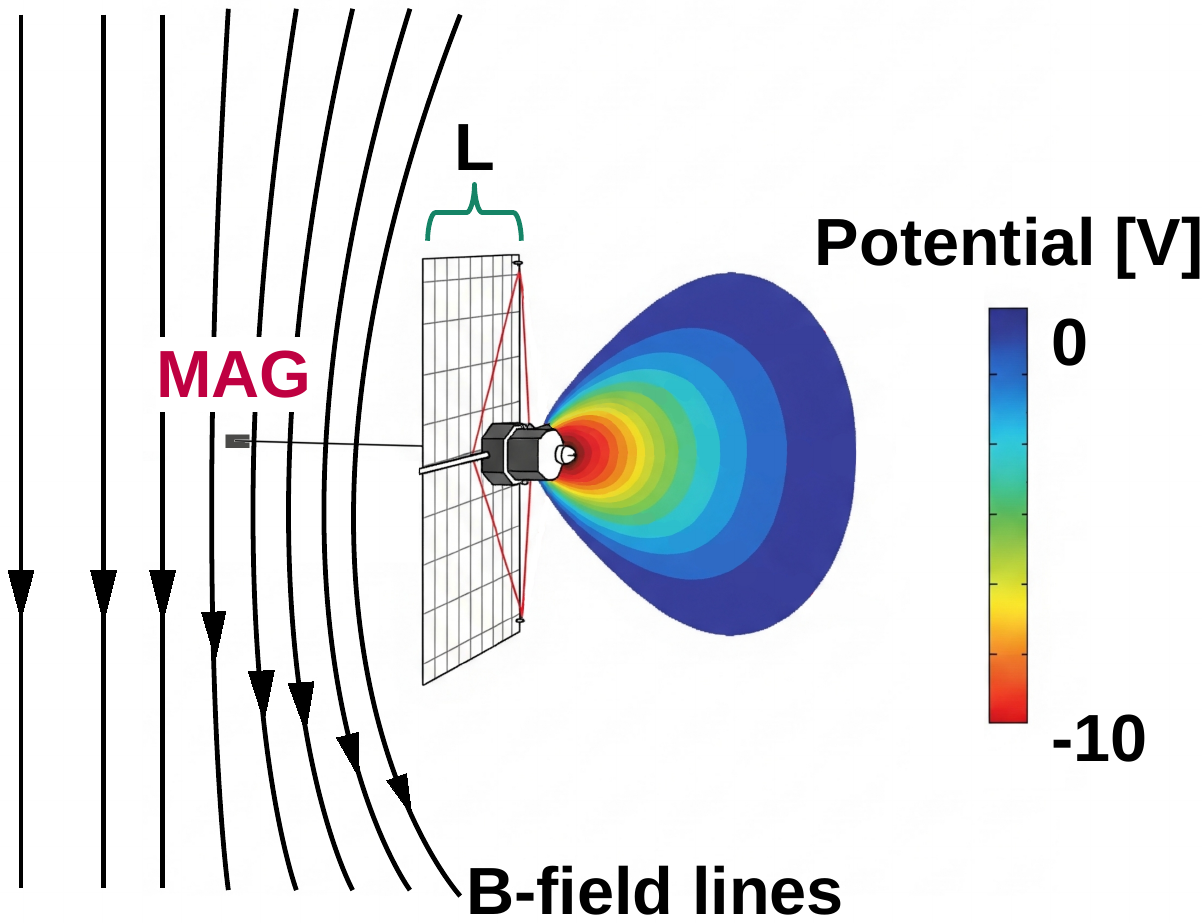}
\caption{Schematic diagram showing the proposed electron-scale magnetic pileup effect. As the electrons approach the negative potential well in the spacecraft wake (shown with an assumed depth of  $k_B T_e/e\sim 10~\text{V}$), they are decelerated as they flow around the barrier. This leads to pileup of the magnetic field, which is frozen into the electron fluid.  The side length $L$ of the solar sail, labeled in the diagram, is assumed to be in the electron fluid regime $\rho_e < L < \rho_p$. A boom-mounted magnetometer (``MAG'') on the sunward side could erroneously sample this disturbed field. {\bf Note on methods:} The schematic diagram portrayed above was created by prompting Google's Gemini generative AI and then modifying the output with conventional image-editing software. }
\label{sail_magnetic_pileup_fig}
\end{figure}

We propose, however, that the negatively-charged electrons may be significantly obstructed if the electrostatic potential $\phi(\bvec{x}, t)$ of the sail or its vicinity becomes sufficiently negative. While the floating potential of conducting satellites and solar sails trend a few volts positive due to significant photoemission of electrons from sunlit surfaces \cite<e.g., >[]{pulupa14, wilson23_wind_sc_potential, parker_2005_solar_sail_electrostatic_charging}, the downstream wake of the spacecraft instead tends to develop a large electrostatic potential well ($\phi < 0$). The magnitude of the wake potential $|e \phi|$ is of the order $k_B T_e \sim$10~eV \cite{engwall_2006b_spacecraft_wake_formation, andre_2021_spacecraft_wake}. This estimate holds for a spacecraft much larger than the Debye length ($L >> \Lambda_D$) \cite{eriksson_2007_sw_spacecraft_wakes}. The fluid-like thermal population of electrons would then be diverted around this potential well, causing magnetic pileup in a region that includes the sunward side of the spacecraft---this scenario is pictured schematically in Figure~\ref{sail_magnetic_pileup_fig}. 

Such a process of electron-scale pileup has been observed in space \cite{kivelson_1993_galileo_at_gaspra_asteroid, zhang_2021_whistler_wings_lunar_magnetic_anomalies}. To our knowledge, the natural occurrence of this phenomenon was first reported in~\cite{kivelson_1993_galileo_at_gaspra_asteroid}, using the Galileo satellite's in situ magnetometer measurements near the asteroid Gaspra. The authors noted that the magnetic field was draped around Gaspra, in a U-shaped pattern consistent with ``whistler wings'', which had before then only been observed in the laboratory \cite{stenzel_1989_whistler_wings_in_lab}. This name derives from the fact that at sub-proton scales, the whistler mode mediates the motion of the electron fluid, including its diversion around an obstacle. Whistler wings can be formed when the plasma moves relative to a magnetic dipole or a pulsed current (i.e. a moving charge density); or, conversely the obstacle can be stationary while the plasma flows by \cite{stenzel_1999_whistlers_in_space_and_laboratory}. The pulsed-current experiments of \cite{stenzel_1989_whistler_wings_in_lab} therefore imply that the net negative charge of the spacecraft's potential wake, which appears stationary in the spacecraft frame, should be capable of producing a draped magnetic field in the moving solar wind electron fluid.




The $\sim$keV solar wind protons (speeds $\sim v_{sw}$) are much more energetic than the sail potential, implying a ``narrow''-type spacecraft wake with transverse width $W\sim L$ \cite<e.g., >[]{engwall_2006b_spacecraft_wake_formation, andre_2021_spacecraft_wake}. This geometry of the wake results from the sail's absorption of ballistically-propagating ions, creating a shadow in the shape of the sail. We note the requirement $L \gtrsim \rho_e$, as stated above, is only a proxy for the more apt requirement $W>\rho_e$ that determines whether thermal electrons will be fluid-like upon encountering the wake.
Simulations are needed, though, to determine the geometry of the wake structure and the electron-scale pileup.

\section{Discussion}\label{discussion_sec}

In Section~\ref{interaction_sec}, we argued that the eddy currents and magnetic pileup are relevant to magnetometer measurements in the presence of a solar sail. The magnetic perturbations caused by these effects are more concerning for larger sails and (in the case of eddy currents) higher-frequency fluctuations. The hypothetical Solar Cruiser-like mission, with a square sail size of $L\sim 40~m \ll \rho_e$ and magnetometer sampling frequency $f_{mag}\sim10$~Hz (Table~\ref{params_table}), is well within the thresholds established here and is unlikely to be compromised by these effects. Another future implication of our study is that eddy currents in a solar sail can significantly perturb high-frequency ($\sim$kHz) magnetometer measurements (Figs.~\ref{sail_induced_B_fig}, \ref{sail_induced_Btot_fig}). This may limit the use of search coil magnetometers on solar sail missions.

Our account of the sources of magnetic perturbations is by no means exhaustive.  Several other magnetic cleanliness issues have been identified in the literature---for example, the thermoelectric effect \cite<e.g., >[]{schnurr19_goes_mag_lessons, acuna04}, photoelectric ``spacecraft charging currents, in response to differential illumination'' as proposed in \cite{eastwood14}, and transient fields from electrostatic arcing. These effects are more difficult to estimate in general, as we anticipate they depend even more on the detailed geometry and material composition of the satellite, and are not covered here in detail.
It is generally recognized that a junction of different conducting materials (with different Seebeck coefficients) is required to generate a current via the thermoelectric effect---so such thermoelectric currents would not depend on the sail alone but rather on the connections between the sail and other parts of the spacecraft. As long as proper magnetic cleanliness practices are followed \cite{acuna04, schnurr19_goes_mag_lessons}, avoiding contact of dissimilar metals with the sail coating, any fields from the thermoelectric effect will be negligible. 
Spacecraft charging currents will be transient, not persistent, as the system relaxes to a state of ``current balance'' \cite{salem01}---for example, following a change in solar illumination as the orientation of the sail is changed for navigation, or when the spacecraft encounters a change in the solar wind electron density.
As discussed earlier, sunlit potentials on solar sails are predicted to be comparable to those observed on conventional spacecraft in the solar wind \cite{pulupa14, wilson23_wind_sc_potential, parker_2005_solar_sail_electrostatic_charging}. 


In Section~\ref{eddy_sec}, we simplified the analysis of eddy currents in a square sail of side length $L$, by instead considering a circular sail of comparable dimensions (radius $R=L/2$). This approach implicitly assumes that the magnetic fields induced in each system will be similar---say, of the same order of magnitude---for a given applied field $B_z(t)$. Further, we note that even a square sail may not consist of a single sheet, as in practice such sails are typically manufactured from four quadrants \cite<e.g., >[]{stohlman2020_solar_sail_folding}. These geometrical considerations will in general modify our quantitative predictions. Nonetheless, we expect that our estimates for the relative magnitude of the perturbation (pictured in Figures~\ref{sail_induced_B_fig},~\ref{sail_induced_Btot_fig}) are relevant to a broad range of sail geometries with characteristic length-scale $R$. 

We also note that when magnetic pileup is active, the draping of the field lines may align them with the 2D plane of the sail.
So, in this case the sail's eddy current effect, which depends on time-varying normal fields, may be suppressed.
But, the draping by itself can significantly affect both the magnitude and direction of the field. 
Incidentally, the pitch-angle distributions of the magnetized particles would be affected.

We intend to study solar sail magnetic pileup in more detail in future work. This complex problem requires kinetic modeling of both the ion and electron particle species, the electromagnetic field, and the spacecraft-plasma interaction. Particle-in-cell codes may be well-suited for this task, as they have been applied to productively model other kinetic-scale space plasmas, such as the environments around comets \cite{deca17_comet_pic, sishtla19_electron_trapping_comet_pic_simulations} and the proposed ``magnetic solar sail'' design \cite{wada19_magnetic_solar_sail_pic}.



\section{Summary and Conclusions}\label{summary_sec}

We here summarize our main results, concerning how a spacecraft's in situ magnetometer measurements are affected by an onboard solar sail:

\begin{itemize}
\item Applied normally-incident fields from the solar wind can diffuse through a solar sail and induce large-scale eddy currents. For an idealized cylindrically symmetric model of the sail, the perturbation field $\bvec{B}^\prime(\bvec{x}, t)$ can be predicted analytically in 3D-space (eq. \ref{B_rings_general_eq}). On the axis of symmetry ($z$), the perturbation field is axial and predicted by eq.~\ref{B_rings_zaxis_eq}.
\item For a sinusoidal applied signal, the eddy-current perturbation field is proportional to both the frequency $f$ and characteristic scale $h \cdot R$ of the sail: $B^\prime_z(z, t)\propto f h R$ (eq.~\ref{eddy_current_B_eq_w_phase}). Search coil magnetometers operating in the $\sim$kHz range are significantly compromised by this effect (Figs.~\ref{sail_induced_B_fig},~\ref{sail_induced_Btot_fig}).
\item  For a larger sail in the regime $\rho_e<L<\rho_p$, magnetic pileup becomes a concern because the magnetic field is frozen in the electron fluid. We propose that pileup could occur when the electrons are decelerated and diverted around negative-potential regions in the sail's vicinity, such as the $\phi(\bvec{x}) <0$ wells that commonly form in spacecraft wakes.
\end{itemize}
For a hypothetical mission, with sail size $L\sim 40~m\ll \rho_e$ and magnetometer frequency range $f\leq f_{mag}\sim 10$~Hz:

\begin{itemize}
\item The relative error ($\mathcal{E}_1$) of the vector magnetic field due to eddy currents is $\lesssim 0.1\%$, satisfying the observational requirement $\Delta B_i/B_i < 0.01$ (Table~\ref{params_table}).
\item No magnetic pileup is expected, because neither protons nor electrons behave as a fluid at such scales.
\end{itemize}

Our results can be used to estimate an upper bound on the sail size $L$, where magnetic pileup can be avoided. Let us adopt the common assumption that the average solar wind density scales as the inverse square of heliocentric distance $\mathcal{R}$. Furthermore, let us assume that for all $\mathcal{R}$ the maximum observed density is a fixed multiple of the average density $n_{max}(\mathcal{R})/n_{ave}(\mathcal{R})\approx 25$ (\cite{liu_2014_stereo_extreme_cme}, see section~\ref{pileup_sec}). The average inertial length at 1~AU is $\rho_e = 2.3\times10^3$~m (Table~\ref{params_table}). Then, from the onset condition of magnetic pileup, $L < \rho_e$, we obtain the allowable $L$, as a function of the minimum heliocentric distance $\mathcal{R}_{min}$ explored by the satellite:

\begin{equation}\label{L_threshold_eq}
\begin{split}
L &< (2.3\times10^3~m)\Big(\frac{\mathcal{R}_{min}}{1~AU}\Big)\sqrt{\frac{n_{ave}}{n_{max}}} \\
& = (4.6\times 10^2~m)\Big(\frac{\mathcal{R}_{min}}{1~AU}\Big).
\end{split}
\end{equation}

\noindent From equation~\ref{L_threshold_eq}, we can see for example that a satellite exploring at a heliocentric distance $\mathcal{R}_{min}=0.5~AU$ should avoid a sail larger than $L=230$~m in order to avoid magnetic pileup measured during the most impactful, high-density space weather events. 

Similarly, we may estimate the range of sail sizes for which eddy currents may be ignored, at all frequencies $f<f_{mag}/2$ within the magnetometer's Nyquist frequency. We require that the relevant error metric $\mathcal{E}$ ($\mathcal{E}_1$ or $\mathcal{E}_2$) should not exceed a given error threshold $\max |\Delta B_i/B_i|$ of the vector field components:  $\mathcal{E}<\max |\Delta B_i/B_i|$. For example, assuming $\mathcal{E} = \mathcal{E}_1$ (equation~\ref{eddy_current_B_err_eq}), we invert the inequality to find another constraint on the sail size $L$. Using parameters given in Table~\ref{params_table}, we find:

\begin{equation}\label{f_mag_threshold_eq}
\begin{split}
L&< \frac{8}{\pi \mu_0 \sigma h f_{mag}}\Big| \frac{\Delta B_i}{B_i}\Big|\\
&= (5.36 \times 10^2~m) \Big(\frac{\sigma_{Al}}{\sigma} \Big)\Big(\frac{10^{-7}m}{h} \Big) \Big(\frac{10~Hz}{f_{mag}}\Big).
\end{split}
\end{equation}


As multiple solar sail missions have successfully proven the feasibility of navigation using only the Sun's photon pressure, this technology is ready for use on a magnetometer-carrying spacecraft.
The magnetic field is one of the fundamental characteristics of space plasmas---and notably the southward component of the field is a key predictor of space weather. 
It is therefore crucial to understand the physics of sail-plasma interaction and how it may impact the onboard field measurements.  
Identifying the ways that solar sails limit observations may also help to inspire new technological developments to mitigate these effects.
Our analysis shows that a solar sail mission is unlikely to suffer from the effects of eddy currents and magnetic pileup. 
The estimates provided in this study may also help to guide the design of future spacecraft and their onboard magnetometers.

\appendix
\section{Eddy Currents in a Circular Sail}\label{appendix_sec}
{
In the presence of a homogeneous time-varying field $\bvec{B}(\bvec{x}, t) = \bvec{\hat z} B_z(t)$, an azimuthal electric field $\bvec{E}(\bvec{x}, t) = \bm{\hat \theta} E_\theta(r, t)$ is induced in a conducting disk  centered at the origin in the $x-y$ plane. From the integral form of Faraday's law, the electric field is $E_\theta(r, t) = -r\dot B_z(t)/2$. Combining this with Ohm's Law $\bvec{J} = \sigma \bvec{E}$ in the conducting material, the azimuthal current $\bvec{J}(\bvec{x},t) = \bm{\hat \theta} J_\theta(r, t)$ in the disk is:


\begin{equation}\label{J_eq}
J_\theta(r, t) = -\frac{\sigma r}{2}\dot B_z(t).
\end{equation}

\noindent We can divide our system into concentric rings with height $h$ and infinitesimal radial breadth $dr$. Then the infinitesimal current $dI_\theta(r, t)$ flowing through each ring is simply the current density $J_\theta(r,t)$ multiplied by the cross-sectional area $h\cdot dr$:

\begin{equation}\label{dI_eq}
dI_\theta(r, t) = -\frac{\sigma r h}{2}\dot B_z(t) dr.
\end{equation}

We may calculate the perturbation field $\bvec{B^\prime}(\bvec{x}, t)$ by summing the contributions from the induced current loops, as pictured in Figure~\ref{tan_fig2}b. To this end, we apply the result from magnetostatics to our (slowly evolving) system, namely that a current loop of radius $a$ in the $x-y$ plane produces a magnetic field $\bvec{B}_{loop}$ proportional to the azimuthal current $I$ that flows through it: $\bvec{B}_{loop}(\bvec{x}) = I \bm{\beta}(\bvec{x}; a)$. Note that we denote the radius of the loop by ``$a$'' instead of ``$r$'', in order to emphasize its role as a parameter rather than a spatial coordinate. The function $\bm{\beta}(\bvec{x}; a)$ describes the spatial variation of the field and is known analytically given an arbitrary radius $a$; see equations 24-25 of \cite{simpson01_current_loop_B_field}. For our system, we can integrate this known relation over the disk (ignoring the finite height $h$) to obtain the total perturbation field:

\begin{equation}\label{B_rings_general_eq}
\bvec{B^\prime}(\bvec{x}, t) = \int_{disk} \bm{\beta}(\bvec{x}; a) dI_\theta(a, t).
\end{equation}

Although the general equations for $\bm{\beta}(\bvec{x}; a)$ given in \cite{simpson01_current_loop_B_field} are relatively difficult to integrate analytically, we can simplify the problem by considering only positions $\bvec{x}$ on the $z$-axis. It is elementary to derive the axial magnetic field of a circular current loop from the Biot-Savart law, and the result may be written in our notation as:

\begin{equation}\label{beta_r0_eq}
\bm{\beta}(\bvec{x}; a)\Big |_{r=0} = \bvec{\hat z}\frac{\mu_0 a^2}{2(a^2 + z^2)^{3/2}}.
\end{equation}

\noindent Combining equations (\ref{dI_eq}), (\ref{B_rings_general_eq}), (\ref{beta_r0_eq}), we finally obtain the axial magnetic field:

\begin{equation}\label{B_rings_zaxis_eq}
\begin{split}
\bvec{B^\prime}(\bvec{x}, t)\Big |_{r=0}
&= \int_{disk} \bm{\beta}(\bvec{x}; a)\Big |_{r=0} dI_\theta(a, t) \\
&= -\bvec{\hat z} \frac{\mu_0 \sigma h}{4} \dot B_z(t)\int_0^R  \frac{a^3}{(a^2 + z^2)^{3/2}} da \\
&= \boxed{-\bvec{\hat z} \frac{\mu_0 \sigma h}{4} \dot B_z(t) \frac{(|z| - \sqrt{R^2 +z^2})^2}{\sqrt{R^2 +z^2}}} = \bvec{\hat z} B^\prime_z(z, t).
\end{split}
\end{equation}

\noindent We note that $B^\prime_z(z,t)$ is even with respect to $z$, and $\partial B_z^\prime/\partial z$ is discontinuous at $z=0$. This latter quality arises from a boundary condition of the vector potential $\bvec{A}(\bvec{x}, t)$ in classical electromagnetism (where $\bvec{B} = \nabla \times \bvec{A}$). Namely, the normal derivative $\pa\bvec{A}/\pa n $ is discontinuous about a surface current with normal unit vector $\bvec{n}$ \cite<>[chapter~5]{griffiths_book_2017_intro_electrodynamics}, and $\pa \bvec{B}/\pa n = \nabla \times (\pa \bvec{A}/\pa n)$ inherits this discontinuity. 

%



%
%

\section*{Open Research}

No original data was used to produce this manuscript.

\section*{Conflict of Interest disclosure}
The authors declare there are no conflicts of interest for this manuscript.


\acknowledgments
This work was supported by NOAA cooperative agreement NA220AR4320151. 
The statements, findings, conclusions, and recommendations are those of the authors and do not necessarily reflect the views of NOAA or the U.S. Department of Commerce.
The authors thank C.~Y.~Tan and D.~Verscharen for permitting the use of previously reported figures. 

%
%

\bibliography{paper_refs}

%
%
%
%
%

\end{document}